\documentclass[prl,twocolumn,amsmath,amssymb,floatfix]{revtex4}
\usepackage{natbib}
\usepackage{dcolumn}
\usepackage{graphicx}
\usepackage{bm}
\usepackage{scrtime}[24hr]
\usepackage[usenames]{color}

\begin{document}

\newcommand{\degc}{\ensuremath{^{\circ}}C}

\bibliographystyle{apsrev}

\title{Measurement of Effective Temperatures in an Aging Colloidal Glass}
\author{Nils Greinert}
\author{Tiffany Wood}
\author{Paul Bartlett}
\email{P.Bartlett@bristol.ac.uk}

\affiliation{School of Chemistry, University of Bristol, Bristol BS8
1TS, UK.}


\begin{abstract}

We study the thermal fluctuations of an optically confined probe
particle, suspended in an aging colloidal suspension, as the
suspension transforms from a viscous liquid into an elastic glass.
The micron-sized bead forms a harmonic oscillator. By monitoring the
equal-time fluctuations of the tracer, at two different laser
powers, we determine the temperature of the oscillator,
$T_{\text{o}}$. In the ergodic liquid the temperatures of the
oscillator and its environment are equal while, in contrast, in a
nonequilibrium glassy phase we find that $T_{\text{o}}$
substantially exceeds the bath temperature.

\end{abstract}

\maketitle

Understanding the slow dynamics of glasses is one of the most
fascinating yet difficult challenges in statistical physics. One
question which has attracted considerable interest is whether the
dynamical fluctuations of a glass can be characterized by a
non-equilibrium temperature. A rigorous thermodynamic temperature is
strictly impossible to define for an aging glass, which remains far
from thermal equilibrium even on long-time scales. Nevertheless
various groups \cite{2087,4260} have proposed ``effective''
temperatures with many of the properties expected for an equilibrium
temperature.  These ideas lead to the surprising prediction of
\textit {two} distinct temperatures in an aging glass \cite{2087}.
The fast rattling modes of particles inside the cage constituted by
neighbors thermalize rapidly to the temperature of the environment,
$T_{\text{bath}}$, while the much slower structural rearrangement of
these cages are supposed to be characterized by a second
temperature, $T_{\text{eff}}$, which mean-field models predict
should exceed $T_{\text{bath}}$. To date most of the support for
this striking two-temperature picture has emerged from simulation
results \cite{4260} on idealized glasses. Experiments have so far
produced conflicting results. Studies of colloidal glasses have
reported that $T_{\text{eff}}$ increases \cite{3631}, remains
unchanged \cite{4166}, or even decreases \cite{1852} with age in
contrast to measurements on structural \cite{1719} and spin glasses
\cite{2728} which have revealed effective temperature warmer than
the bath temperature.

In this Letter, we report the temperature of a micrometer-sized
sphere immersed in an aging colloidal suspension, as the suspension
transforms from a fluid to a glass. The particle, captured in an
optical trap, constitutes a microscopic harmonic oscillator whose
fluctuations probe the nonequilibrium dynamics of the aging glass.
We measure the equal-time fluctuations of this local oscillator and
show that, in an ergodic phase, the temperature of the oscillator
$T_{\text{o}}$ equals the environment temperature $T_{\text{bath}}$
of the system, as required by equilibrium statistical mechanics.
Significantly, when we repeat the measurements in an aging glass we
find a higher temperature and $T_{\text{o}}
> T_{\text{bath}}$.

All experiments to date on colloidal glasses \cite{3631,4166,1852}
have relied on active, driven measurements. We conduct our
experiments instead in a quasi-static limit which has several
advantages. First, the experiments are simpler because there is no
need to characterize the complete time-dependent response of the
system. Second, the lack of an external driving force ensures that
we always operate within the linear response regime. We determine
the temperature of the laser-trapped particle by recalling that the
fluctuations in the coordinate, $\delta x = x - \left <x \right
>$, of an oscillator (with spring constant $k_{\text{T}}$), in contact with an equilibrium system at a
temperature $T$, are Gaussian,
\begin{equation}\label{pdf}
 P(\delta x) = \frac{1}{\sqrt{2 \pi T \chi}} \exp \left [
   - \frac{(\delta x)^{2}}{2 T \chi}  \right ].
\end{equation}
Here $k_{B} = 1$ and the static susceptibility $\chi$ -- the shift
in the mean position $\left < x \right >$ produced by a small
constant external field -- is simply $1/k_{\text{T}}$.
Operationally, we define the temperature of the oscillator,
$T_{\text{o}}$, by the ratio between the one-dimensional
mean-squared displacement (MSD) of the probe and the static
susceptibility, $ T_{\text{o}} = \left < \delta x^{2} \right
> / \chi$. Equilibrium statistical mechanics guarantees $T_{\text{o}} = T$,
irrespective of the nature of the coupling between oscillator and
system. Out-of-equilibrium however this equivalence no longer holds.
Berthier and Barrat \cite{2752} have argued that a generalized
equipartition principle holds in nonequilibrium glassy materials,
implying that an oscillator will record the (slow) effective
temperature of a glass. We postpone the interpretation of
$T_{\text{o}}$ until later and use it, for now, simply as a
experimentally accessible measure of the local temperature of a
material.

Brownian fluctuations change considerably at the glass transition
simply because of elasticity. A fluctuating
optically-trapped sphere in an elastic medium, such as a glass, is
subject to two harmonic forces; an optical force $F_{\text{opt}}$,
resulting from the external laser field, and an additional elastic
force $F_{\text{e}}$, caused by matrix deformation. In an infinite
continuum, with shear modulus $G$ and Poisson ratio $\nu$, the
elastic force is a linear function of position, $F_{\text{e}} =
-k_{\text{e}} (x - \left <x \right
>)$. The spring constant is
 $k_{\text{e}} = 6 \pi G R$ for a sphere of radius $R$, when $\nu = 1/2$ \cite{1833}.
The optical forces are $F_{\text{opt}} = -k (x - \left < x \right
>)$, with $k$ the optical trap stiffness, so the
total spring constant of the trapped particle is $k_{\text{T}} =
k_{\text e} + k$. To separate the elastic contribution from the
optical contributions to $k_{\text{T}}$ we modulate the intensity of
the laser beam, rapidly switching it between two levels, which we
identify below by the subscripts 1 and 2 respectively. Since
$k_{\text{e}}$ is independent of laser power, consecutive
measurements of $\left < \delta x^{2} \right> $ at two different
laser powers  provides estimates of both $k_{\text{e}}$ and
$T_{\text{o}}$,
\begin{eqnarray} \label{eq:twopower}
  k_{\text{e}} & = & \frac{k_{2} \left < \delta x^{2} \right >_{2} - k_{1} \left < \delta x^{2} \right >_{1} }
  {\left < \delta x^{2} \right >_{1} - \left < \delta x^{2} \right >_{2}}  \nonumber \\
  T_{\text{o}} & = &  \frac{(k_{2} - k_{1}) \left < \delta x^{2} \right >_{1} \left < \delta x^{2} \right >_{2} }
  {\left < \delta x^{2} \right >_{1}- \left < \delta x^{2} \right
  >_{2}}.
\end{eqnarray}
Here $k_{i}$ ($i = $ 1, 2) is the optical force constant at each
laser power.

The glass studied was a transparent aqueous dispersion of charged
colloidal disks of radius $a \approx 15$ nm and 1 nm thick (Laponite
RD).  The suspension (2.4\% mass fraction) was filtered through a
0.45 $\mu$m filter to obtain a reproducible initial liquid at $t =
0$. Immediately after filtration, a small amount of a dilute
suspension of silica spheres (radius $R$ = 0.55 $\pm$ 0.03 $\mu$m)
was added. A particle was captured in a tightly-focused laser beam
($\lambda = 1064$ nm) and its lateral position, $x(t)$, measured
with a quadrant photodetector. The strength of the optical trap was
cycled every 52 s between an initial (optical) trap strength, $k_{1}
= 4.4$ pN/$\mu$m, and a final trap strength, $k_{2} = 11.0$
pN/$\mu$m. All experiments were performed at 22 $\pm $ 1 \degc.  We
confirmed that laser heating was negligible by following the
temperature of a trapped silica bead in deionized water. Recording
continuously for over an hour at ambient conditions gave a mean
temperature of $T_{\text{o}} = 300 \pm \; 8$ K \cite{note2}, with no
upward drift in $T_{\text{o}}$.  Repeating the measurements on
\textit{different} particles showed that the small variation in bead
size led to a systematic error in $T_{\text{o}}$ of about $\pm \;
40$ K.

\begin{figure}[h]
\includegraphics[width=2.8in]{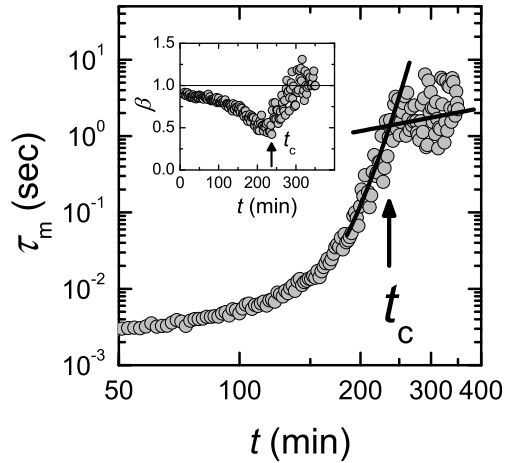}
\caption{Variation of the mean relaxation time $\tau_{\rm{m}}$ with
time after preparation for $k_{2} = 11.0 $ pN/ $\mu$m. The
correlation function $C_{\text{x}}(\tau)$ was fit to a
stretched/compressed-exponential relaxation. The continuous lines
are
 fits to the exponential and linear aging regimes
 respectively.  $t_{c}$ identifies the crossover from exponential to the full aging
 regime and indicates the approximate position of dynamical arrest. Similar plots are found if the relaxation
 time is chosen as $\tau_{\alpha}$ or defined as the time at which
 $C_{\text{x}}(\tau)$ has decayed by a factor of $1/e$.
 Inset: The age dependence of the fitted exponent $\beta$. } \label{fig:relaxtion}
\end{figure}

After mixing the suspension gradually thickens with
time and becomes jammed in a glassy state \cite{2206}. To identify
the point of dynamical arrest we measure the autocorrelation
function of the Brownian fluctuations of the probe sphere,
$C_{\text{x}}(\tau) = \left < \delta x(\tau) \delta x(0) \right > /
\left < \delta x^{2} \right
> $. The positional fluctuations are well represented by the
stretched exponential relaxation, $C_{\text{x}}(\tau) = \exp \left [
- (\tau/\tau_{\alpha})^{\beta} \right ] $, familiar from generic
glassy systems, where $\tau_{\alpha}$ is a terminal relaxation time
and $\beta$ is a stretching exponent. The inset to
Fig.~\ref{fig:relaxtion} reveals that the fitted exponent $\beta$
varies considerably with age. For $t < t_{c}\; (\approx 220 \pm 15 $
min) the correlation function is stretched ($\beta < 1$) but with
increasing age there is a crossover from a stretched- to a
near-exponential form. To decouple changes in the relaxation time
from the changes in the exponent $\beta$ we evaluate the `mean'
relaxation time, $\tau_{\rm{m}} = (\tau_{\alpha}/\beta) \Gamma(
\beta^{-1}) $ where $\Gamma$ is the gamma function. The relaxation
dynamics clearly exhibit two distinct regimes. There is first a fast
regime where $\tau_{\rm{m}}$ grows approximately exponentially with
sample age, followed by a second regime of much slower growth where,
$\tau_{\rm{m}} \sim t$. Dynamical arrest occurs at a time $t_{c}$
which we estimate from the intersection of the corresponding fits in
Fig.~\ref{fig:relaxtion} as 220$\; \pm \; 15$ min. For $t < t_{c}$,
the system behaves as if it were an ergodic liquid, albeit one with
a relaxation time which slows down progressively as $t$ increases.
At $t \approx t_{c}$ the suspension arrests and begins to age.

\begin{figure}[h]
\center{
\includegraphics[width=2.8in]{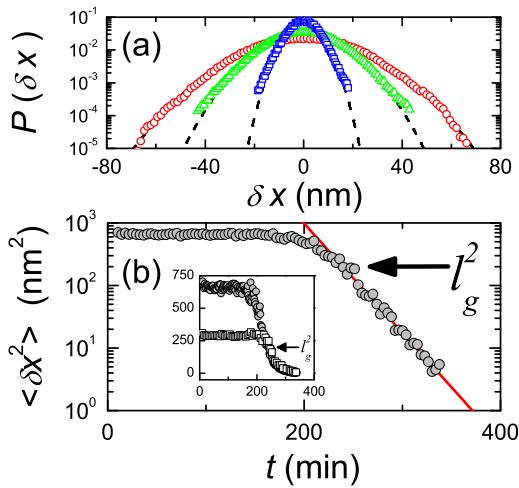}
\caption{(color online). (a) Probability distributions of the
fluctuations of probe particles for different times after
preparation, $t = 200$ min (circles), 250 min (triangles), and 301
min (squares). The dashed lines are Gaussian fits to the data. (b)
Asymptotic mean-square displacement (MSD) of probe, confined in an
optical trap of strength $k_{1} = 4.4 $ pN/$\mu$m, as a function of
sample age $t$. The statistical errors are of order $\pm 11$
nm$^{2}$. The arrow indicates the square of the localization length
in the glass, $l_{g}^{2}$. The solid line is an exponential fit to
the MSD within the glass. The inset depicts a linear plot of the MSD
(in nm$^{2}$) as function of age, in weak ($k_{1}$: circles) and
strong ($k_{2}$: squares) laser traps. $k_{2} = 11.0 $ pN/$\mu$m.
Note the  MSDs are identical, within error, for $t \agt 250$ min
where $\left < \delta x^{2} \right
> \alt l_{g}^{2}$.}
 \label{fig:gaussians}
 }
\end{figure}

To explore the effect of the glass transition on the temperature
recorded by the Brownian probe, we first examine the probability
distribution of the fluctuations, $P(\delta x)$.
Fig.~\ref{fig:gaussians}(a) reveals that although the particle is
highly constrained by the glassy matrix, the trajectories exhibit
fluctuations which remain Gaussian. This implies that
the Einstein
 expression for the fluctuations (Eq.~\ref{pdf}) can be generalized by replacing the
 bath temperature with an effective temperature. To measure the
 effective temperature we determine the MSD of the trapped particle, recognizing that the dramatic dynamical slowing at the glass transition, evident in
Fig~\ref{fig:relaxtion}, requires care if the asymptotic limit is to
be evaluated. To check for convergence, we split each
recorded 52 s particle trajectory into a number of shorter
equal-time lengths of duration $t_{B}$ and evaluate the MSD of each,
before averaging the values together. We choose $t_{B}= 3.3$ s in
order to have a good signal to noise ratio whilst retaining many
long-lived fluctuations. The elastic modulus $G$ of a (hard-sphere)
repulsive glass scales as $G \sim k_{B}T / a^{3}$, where $a$ is the
radius of the particles that form the glass, so the thermal
fluctuations of a sphere of radius $R$ embedded in a glass are
restricted to a cage with a characteristic size of $l_{g} \sim (
a^{3} / R)^{1/2}$ \cite{1833,Mason-664}. Taking the effective radius
as $a \approx 45$ nm \cite{note} yields $l_{g}^{2} \approx 200$
nm$^{2}$ for our system. Consistent with this picture the asymptotic
MSD exhibits two distinct regimes. Fig.~\ref{fig:gaussians}(b)
demonstrates that for $t \alt 100$ min, where $\left < \delta x^{2}
\right >
> l_{g}^{2}$, the degree of localization does not change with age although it depends
sensitively on the laser intensity, as expected for confinement by
an external field. With increasing age, by contrast, the asymptotic
MSD shows a progressively weaker dependence on intensity. Indeed
when $\left < \delta x^{2} \right
> \approx l_{g}^{2}$ the thermal fluctuations are almost totally
independent of the laser field, consistent with permanent caging of
the probe particle by the glass.  The localization length in the
glass shrinks exponentially as the glass get older, until at very
long times ($t \simeq 350$ min) all measurable motion ceases.


\begin{figure}[h]
\includegraphics[width=2.6in]{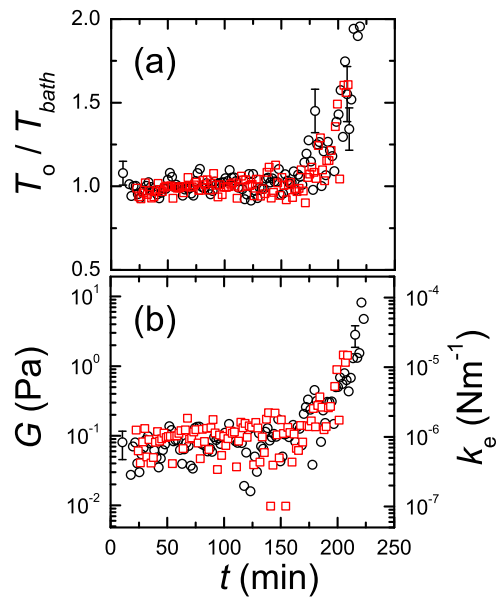}
\caption{(color online) (a) Effective temperature as a function of
sample age. The circles and squares depict the results of
measurements using similarly-sized particles in two
physically-distinct laponite samples. The scatter illustrates the
typical level of reproducibility achieved. Errors bars, which are
estimated from the variability in $\left < \delta x^{2} \right >$
for $t < 100$ min where the MSD is constant, are only shown on a few
representative points for clarity. (b) Age dependence of the single
particle elastic spring constant $k_{\text{e}}$ and the equivalent
elastic modulus $G = k_{\text{e}} / 6 \pi R $.}
\label{fig:elasticity}
\end{figure}

The change in localization with time, evident in
Fig.~\ref{fig:gaussians}(b), allows us to monitor the temperature of
the slow structural modes of the glass. Figure~\ref{fig:elasticity}
shows the age dependence of the oscillator temperature
$T_{\text{o}}$ and the shear modulus $G$, calculated from
Eq.~\ref{eq:twopower} and the data of Fig.~\ref{fig:gaussians}(b)
(assuming $\nu = 0.5$).  The data reveals a marked increase in
$T_{\text{o}}$ approaching the liquid-glass transition. For $t \ll
t_{c}$ the temperature of the Brownian probe is approximately
constant at $247 \pm \; 20$ K \cite{note2}, which given that the
systematic uncertainty in $T_{\text{o}}$ is at least $40$K, is in
reasonable agreement with the environment temperature,
$T_{\text{bath}} = 295$ K. By contrast, in the glass  we find a
systematic increase in the oscillator temperature with increasing
age, with the ratio $T_{\text{o}}/T_{\text{bath}} \approx 2$ at the
longest accessible age. Since the glass is non-ergodic an ensemble
average quantity such as the effective temperature strictly requires
an average over many probe particles. To verify the reproducibility
of our results we repeated measurements using similarly-sized
particles in different physical samples with the same nominal
laponite concentration and found no significant difference in the
age dependence of $T_{\text{o}}/T_{\text{bath}}$. An example of the
data spread for data sets recorded from two different samples is
shown in Fig.~\ref{fig:elasticity}(a). We attribute this homogeneity
to the fact that the probe is significantly larger than the laponite
particles.


\begin{figure}[h]
\includegraphics[width=2.8in]{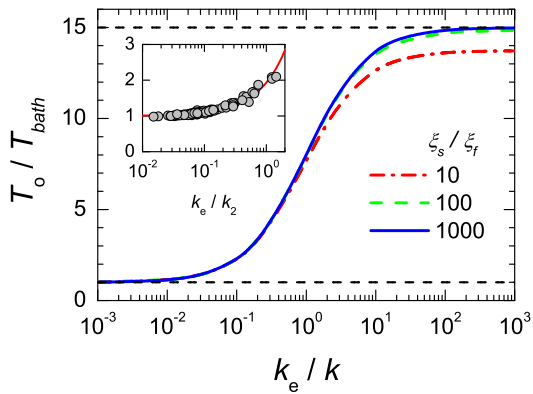}
\caption{(color online) Elasticity dependence of the scaled
oscillator temperature for $T_{\text{eff}} = 15 T_{\text{bath}}$.
The solid lines are obtained from the analytic solution of
Eq.~\ref{eqn:GLE} for relative timescales  $\xi_{s} / \xi_{f} = 10$
(red), 100 (green), and 1000 (blue). Inset: Experimental dependence
of $T_{\text{o}} / T_{\text{bath}}$ on the elasticity ratio,
$k_{\text{e}} /k_{2}$ (circles). The continuous line depict the
theoretical predictions for $\xi_{s} / \xi_{f} = 100$.}
\label{fig:Barrat}
\end{figure}

The growth evident in $T_{\text{o}}$ as the glass ages is, at first
sight, very surprising. Intuitively, one expects a glass to cool as
it ages and $T_{\text{eff}}$ to fall with increasing $t$. To account
for the temperature increase, we analyze a simple stochastic model
of our experiments, ignoring the detailed microhydrodynamic coupling
between the fluctuations of the embedded probe and the sea of
platelets. The model, introduced in \cite{4189}, considers the
diffusion of a harmonically-bound particle coupled to two thermal
baths. The two baths are kept at different temperatures to mimic the
effects of the \textit{fast} and \textit{slow} modes of the glass on
probe diffusion. The fast bath is held at a temperature
$T_{\text{bath}}$ and exerts an instantaneous friction of
$\xi_{f}\dot{x}(t)$ on the particle while the second, slower bath is
maintained at a (higher) temperature $T_{\text{eff}}$ and is
associated with the memory function, $\Gamma (t)$. In the overdamped
limit, where acceleration can be neglected, Brownian diffusion is
governed by the generalized Langevin equation,
\begin{equation}\label{eqn:GLE}
    k x(t) = - \int_{0}^{t} d\tau \Gamma (t-\tau)\dot{x}(\tau) -
    \xi_{f}\dot{x}(t) + \theta(t)+\gamma(t).
\end{equation}
The random forces due to the fast bath are Gaussian with $\left <
\gamma(t) \right > = 0$ and $\left < \gamma(t) \gamma(\tau)  \right
> = 2\xi_{f} T_{\text{bath}}  \delta(t-\tau)$, whereas the slow bath is
characterized by $\left < \theta(t) \theta(\tau)  \right > =
T_{\text{eff}} \Gamma(t-\tau)$. Approximating the glass by a Maxwell
fluid, with zero shear viscosity $\eta_{s}$ and shear modulus $G$,
the retarded friction on a sphere of radius $R$ is
 $\Gamma(t) = k_{\text{e}} \exp ( -k_{\text{e}} t /
\xi_{s})$, with $k_{\text{e}} = 6 \pi G R$ and $\xi_{s} = 6 \pi
\eta_{s} R$ \cite{1742}. Following Ref.~\cite{4189}, an analytic
expression may be written down for the reduced oscillator
temperature $\widetilde{T} = T_{\text{o}} / T_{\text{bath}}$.
Fig.~\ref{fig:Barrat} shows the dependence of the oscillator
temperature on the elasticity ratio $k_{\text{e}} / k$. Clearly
while the oscillator temperature always lies between the
temperatures of the fast and slow baths, the exact value is sharply
dependent on $k_{\text{e}} / k$. Only for $k_{\text{e}}
\gg k$ does $T_{\text{o}}$ approach the temperature of the slow
bath. Plotting the experimental data for $\widetilde{T}$ as a
function of $k_{\text{e}} / k_{2}$ reveals a very similar trend
(shown in the inset of Fig.~\ref{fig:Barrat}). Consequently we
attribute the increase in the measured temperature with age to the
growing strength of the mechanical coupling between particle and
glass.

In summary, we have measured the quasi-static fluctuations of an
optically-bound probe particle immersed in an aging glass. Using a
generalized equipartition principle we determine the temperature of
the oscillator formed by the trapped probe. In a glass we find an
oscillator temperature which is substantially higher than the
temperature of the environment; while in the fluid the temperatures
of the oscillator and environment are essentially equal. We propose
a simple theoretical model for these nonequilibrium experiments
which reveals that, at high elasticities, the probe thermalizes to
the effective temperature of the slow modes of the glass. Our
findings agree broadly with the results obtained recently by Abou
{\it et al.} and Strachan {\it et al.} \cite{3631}, who used
different techniques to measure $T_{\rm{eff}}$. Taken together this
agreement provides strong experimental confirmation for the
existence of an elevated effective temperature in a glass. Future
work will look at using comparably-sized probe and glass particles
to study the role of dynamical heterogeneities on the effective
temperature.


\begin{thebibliography}{21}
\bibitem{note} Estimated by assuming the effective hard sphere volume fraction at the glass transition is $\phi_{g} = 0.58$.
\end{thebibliography}



\end{document}